\documentclass[prl,aps,showpacs,groupedaddress,longbibliography,superscriptaddress,twocolumn,toc=flat,biblatex,footinbib]{revtex4-2}
\usepackage{natbib}
\usepackage[table]{xcolor}
\usepackage[colorlinks=false]{hyperref}
\usepackage[utf8]{inputenc}
\usepackage{color}
\usepackage{bbm} 
\usepackage[english]{babel}
\usepackage{multirow}
\usepackage{amsfonts,amsmath,amssymb,stmaryrd}
\usepackage{braket}
\usepackage{graphicx}

\usepackage{epsfig}
\usepackage{mathrsfs}
\usepackage{verbatim}
\usepackage{centernot}
\usepackage{ulem}
\usepackage{longtable}
\usepackage{nicefrac}
\usepackage[framemethod=tikz]{mdframed}
\usepackage{siunitx}
\usepackage[nolist]{acronym}
\newcommand{\blue}{\color{black}}

\newcommand{\mw}[1]{\textcolor{black}{#1}}

\usepackage{soul}

\usepackage{array}

\usepackage{cancel,ifthen}
\newcommand{\cmnt}[2][NoInPuT]{\ifthenelse{\equal{#1}{NoInPuT}}{}{{\color{red}\sout{#1}}} {\color{blue} #2}}

\usepackage{bm}	
\renewcommand{\vec}[1]{{\bm{#1}}}

\LTcapwidth=\textwidth

\begin{document}

\normalem	

\title{Spurious spin nutation modes arising from truncated memory effects}

\author{Markus Wei{\ss}enhofer}
\email[]{markus.weissenhofer@physics.uu.se}
 \affiliation{Department of Physics and Astronomy, Uppsala University, P. O. Box 516, S-751 20 Uppsala, Sweden}

\author{Ritwik Mondal}
\affiliation{Department of Physics, Indian Institute of Technology (ISM) Dhanbad, IN-826004, Dhanbad, India}

\author{M. S. Mrudul}
 \affiliation{Department of Physics and Astronomy, Uppsala University, P. O. Box 516, S-751 20 Uppsala, Sweden}

\author{Peter M. Oppeneer}
 \affiliation{Department of Physics and Astronomy, Uppsala University, P. O. Box 516, S-751 20 Uppsala, Sweden}

\begin{abstract}
Spin dynamics is conventionally treated as inertia-free. The \ac{iLLG} extends this with a second-order time derivative, predicting a high-frequency nutational mode that recent experiments have claimed to observe and interpreted as direct evidence for spin inertia. We show that when spin inertia is generated by eliminating additional dynamical degrees of freedom, the resulting exact spin-only description is generally non-Markovian. The \ac{iLLG} is recovered as the low-frequency, second-order truncation of the underlying memory kernel. Using a minimal spin--cavity model, we show that the exact eigenfrequencies and those obtained via the \ac{iLLG} differ and that the spin nutation mode predicted by the latter is a truncation-induced high-frequency artifact. This raises a broader question {\blue of the applicability of the} \ac{iLLG} \mw{obtained from} analogous low-frequency truncations, which may likewise generate spurious high-frequency roots. Consequently, agreement of a high-frequency spectral feature with the \ac{iLLG} nutation frequency does not, by itself, establish a physical nutation mode.
\end{abstract}

\maketitle

\begin{acronym}
\acro{iLLG}{inertial Landau-Lifshitz-Gilbert equation}
\acro{nM-LLG}{non-Markovian Landau-Lifshitz-Gilbert equation}
\acro{FDT}{fluctuation dissipation theorem}
\acro{EOM}{equation of motion}
\end{acronym}

\textit{Introduction}---Research in ultrafast magnetism increasingly probes spin dynamics on picosecond and sub-picosecond timescales, where higher-order corrections to the \acl{EOM} may become relevant. In recent years, it has been proposed that the spin degree of freedom itself can acquire an inertial term at such timescales~\cite{Mondal2023}, giving rise to the \ac{iLLG}~\cite{Ciornei2011,Li2015},
\begin{align}
\begin{split}
\label{eq:iLLG}
\dot{\vec{S}}
&=
-\vec{S}\times\vec{H}^{\mathrm{eff}}
+
\alpha\vec{S}\times\dot{\vec{S}}
+
\eta\vec{S}\times\ddot{\vec{S}},
\end{split}
\end{align}
where $\vec{H}^{\mathrm{eff}}$ is the effective field, $\alpha$ the Gilbert damping parameter, and the last term, proportional to the inertia parameter $\eta$, introduces a second-order time derivative into the spin dynamics.

The microscopic origin of this inertial term, however, remains under active investigation. Mondal \textit{et al.} derived an inertial contribution from the relativistic Dirac--Kohn--Sham equation~\cite{Mondal2017}. Other approaches have shown that inertial dynamics can emerge through the coupling of localized spins to additional dynamical degrees of freedom, including fermionic~\cite{Bhattacharjee2012,Kikuchi2015,Sayad2016,Bajpai2019,Reyes-Osorio2025,Faehnle2011,Thonig2017,Moussa2026} and bosonic modes~\cite{Anders2022,Quarenta2024}.

The inclusion of the second-order time derivative increases the order of the \acl{EOM} and thereby introduces an additional dynamical mode. In addition to the conventional precessional mode, the \ac{iLLG} produces a second mode at substantially higher frequencies, set by the inverse inertia parameter $\eta^{-1}$, known as the \textit{spin nutational} mode~\cite{Ciornei2011,Wegrowe2012,Olive2012,Mondal2021,Cherkasskii2021,Mondal2022,Titov2022}.

Importantly, experimental signals have recently been attributed to such nutational modes~\cite{Neeraj2021,Unikandanunni2022,De2025} and interpreted as evidence for inertial spin dynamics. Fitting the observed nutation frequencies within the \ac{iLLG} framework has yielded inertia parameters of a few hundred femtoseconds for metallic ferromagnets including CoFeB, NiFe, and Co~\cite{Neeraj2021,Unikandanunni2022}.

Whether such high-frequency features reflect a genuine nutational mode depends {\blue however} on how the inertial term arises. A second-order equation for a reduced variable can emerge when one dynamical variable is eliminated from a coupled set of first-order equations. A simple example is provided by Hamiltonian mechanics: for $\mathcal{H}(p,q)=\frac{p^2}{2m}+V(q)$, Hamilton's equations yield $\dot{q}=p/m$ and $\dot{p}=-\partial V/\partial q$, and eliminating $p$ gives $m\ddot{q}+\partial V/\partial q=0$, i.e., Newton's \acl{EOM}~\cite{Landau1982}. This reduction is exact because $p$ is related instantaneously to $\dot q$. In contrast, eliminating a dynamical degree of freedom generally produces a non-Markovian memory kernel, {\blue i.e., wherein the state of the system depends on its entire past,} rather than an exact finite-order equation.

In this Letter, we show that spin inertia generated by coupling to additional dynamical degrees of freedom does not, in general, imply a physical nutational mode. The exact spin-only dynamics is then non-Markovian, while the \ac{iLLG} provides only its low-frequency, second-order expansion. Using a minimal spin--cavity model that is exactly solvable in the linearized limit, we demonstrate that this truncation can generate a spurious high-frequency root with no corresponding pole in the full response. We further propose a coupling-strength test to distinguish such a truncation-induced root from a genuine hybridized mode.

\textit{Inertial versus non-Markovian spin dynamics}---We adopt a classical description of spin dynamics in which a localized atomic spin moment is represented by a spin vector $\vec{S}$ of fixed magnitude, $|\vec{S}|=1$~\cite{Eriksson2017}. Its components obey the Poisson algebra $\{S^\alpha,S^\beta\}= \sum_{\gamma=1}^3\varepsilon^{\alpha\beta\gamma}S^\gamma$~\cite{Dzyaloshinskii1980}, which ensures the conservation of the spin length.

For a Hamiltonian $\mathcal{H}=\mathcal{H}(\vec{S})$, the spin dynamics is governed by the first-order equation $\dot{\vec{S}}=\{\vec{S},\mathcal{H}\}=-\vec{S}\times\vec{H}^{\mathrm{eff}}$, where $\vec{H}^{\mathrm{eff}}=-\partial\mathcal{H}/\partial\vec{S}$ denotes the effective field. Including Gilbert damping, proportional to $\vec{S}\times\dot{\vec{S}}$ with dimensionless damping parameter $\alpha$~\cite{Gilbert2004}, extends the Hamiltonian dynamics to the Landau--Lifshitz--Gilbert equation, $\dot{\vec{S}}=-\vec{S}\times\vec{H}^{\mathrm{eff}}+\alpha\vec{S}\times\dot{\vec{S}}$. The Gilbert damping term accounts for the transfer of energy from the spin system to its environment and thereby introduces dissipation. The Landau--Lifshitz--Gilbert equation -- and its stochastic extension~\cite{Brown1963} --  has proved remarkably successful in describing a broad range of magnetic phenomena~\cite{Nowak2007,Eriksson2017,Evans2020}.

{\blue At this point it is informative to mention that}
\textit{effective} inertia terms also emerge in magnetic dynamics by eliminating part of the spin degrees of freedom. {\blue For example,} the \acl{EOM} for the N\'{e}el vector in antiferromagnets~\cite{Baltz2018,Kimel2026} and the collective-coordinate equation for magnetic domain walls~\cite{Doering1948,Weissenhofer2022DW} both acquire second-order time derivatives this way, while the underlying spin dynamics itself remains first order. The spin inertia of Eq.~\eqref{eq:iLLG} is {\blue thus} fundamentally different: it modifies the spin's own \acl{EOM}, not that of a collective coordinate derived from it.

To illustrate the emergence of spin inertia upon eliminating additional degrees of freedom, we consider a spin degree of freedom $\vec{S}$ coupled to additional dynamical variables, collectively denoted by $\vec{x}$. A Hamiltonian $\mathcal{H}=\mathcal{H}(\vec{S},\vec{x})$ gives rise to a set of coupled equations of motion, $\dot{\vec{S}}(t)=F_{\vec{S}}[\vec{S}(t),\vec{x}(t)]$ and $\dot{\vec{x}}(t)=F_{\vec{x}}[\vec{S}(t),\vec{x}(t)]$, where, in the Hamiltonian case, $F_{\mathcal{O}}[\ldots]=\{\mathcal{O},\mathcal{H}\}$.

Formally solving the \acl{EOM} for $\vec{x}(t)$ for a given spin trajectory $\vec{S}(t)$, and substituting the solution back into the \acl{EOM} for $\vec{S}(t)$, exactly transforms the coupled dynamics into a spin-only description that can be written as
\begin{align}
    \begin{split}
    \label{eq:nMsD}
    \dot{\vec{S}}(t)
    &=
     -
     \vec{S}(t)
    \times
     \bigg[
    \vec{H}^\mathrm{eff}(t)
    -
    \int_{-\infty}^\infty
    \hspace*{-0.5em}
    \mathrm{d}t'\
    \theta(t-t')
    \mathcal{K}(t,t')
    \vec{S}(t')
    \bigg].
    \end{split}
\end{align}
This integro-differential equation, which we call the \ac{nM-LLG}, is exact: it reproduces the full dynamics of the original coupled system without approximation. All information about the eliminated degrees of freedom $\vec{x}$ is encoded in the memory kernel $\mathcal{K}(t,t')$, through which the spin at time $t$ is coupled to its entire past; the Heaviside step function $\theta(t-t')$ ensures causality.

Physically, the memory kernel captures the backaction of the eliminated degrees of freedom: energy transferred from the spin into $\vec{x}$ can be stored and later returned, producing retardation effects absent in the Markovian limit. Such non-Markovian spin dynamics arise naturally from integrating out environmental degrees of freedom~\cite{Breuer2016} and have been derived explicitly for spins coupled to classical~\cite{Rueckriegel2015} and quantum-mechanical~\cite{Bajpai2019,Anders2022} environments.

The local-in-time \ac{iLLG} can be obtained by expanding the memory term in time derivatives of the spin around the present time~\cite{Anders2022},
\begin{align}
    \begin{split}
    \label{eq:nM_expansion}
    \int_{-\infty}^\infty
    \mathrm{d}t'\
    \theta(t-t')
    \mathcal{K}(t,t')
    \vec{S}(t')
    =
    \sum_{m=0}^\infty
    \kappa^{(m)}(t)
    \partial_t^m
    \vec{S}(t),
    \\
    \kappa^{(m)}(t)
    =
    \frac{(-1)^m}{m!}    
    \int_{0}^{\infty}
    \mathrm{d}\tau\
    \tau^m
    \mathcal{K}(t,t-\tau).
    \end{split}
\end{align}

The coefficient $\kappa^{(0)}(t)$ renormalizes the static effective field, while $\kappa^{(1)}(t)$ and $\kappa^{(2)}(t)$ generate, respectively, Gilbert-like damping and an inertial contribution, corresponding to time-dependent parameters $\alpha(t)$ and $\eta(t)$. In equilibrium or stationary states, time-translation invariance implies $\mathcal{K}(t,t')=\mathcal{K}(t-t')$, such that the coefficients $\kappa^{(m)}$  become time-independent constants~\cite{Reyes-Osorio2025}, and truncating the expansion at second order recovers the \ac{iLLG}.

\textit{Dynamics of a single spin in a cavity}---To illustrate the implications and potential limitations of truncating the derivative expansion, we consider a minimal model in which a single magnetic moment is coupled to bosonic modes of an electromagnetic cavity. We represent the cavity modes as classical harmonic oscillators with frequencies $\omega_n$ and coordinates $(q_n,p_n)$, linearly coupled to the spin,
\begin{align}
\label{eq:H_toymodel}
\mathcal{H}=H_0(\vec{S}) + \frac{1}{2}\sum_n \left( p_n^2 + \omega_n^2 q_n^2\right) - \sum_n q_n \vec{g}_n \cdot\vec{S}.
\end{align}

This spin-plus-oscillator structure is analogous to the prototypical Caldeira--Leggett model of a system coupled to a bosonic environment~\cite{Caldeira1983a,Caldeira1983b}. In the absence of losses, the cavity-mode equations of motion are
\begin{align}
\ddot q_n+\omega_n^2q_n
=
\vec g_n\cdot\vec S .
\end{align}
To account phenomenologically for cavity losses, one may subsequently include a damping term $\gamma_n\dot q_n$.

The retarded Green's function of the damped harmonic oscillator satisfies $(\partial_t^2 + \gamma_n \partial_t +\omega^2_n)G_n(t-t')=\delta(t-t')$ and, in the underdamped regime, is given by
\begin{align}
    G_n(t-t')=\theta(t-t')\frac{\sin[\bar{\omega}_n(t-t')]}{\bar{\omega}_n} e^{-\frac{\gamma_n(t-t')}{2}},
\end{align}
with $\bar{\omega}_n=\sqrt{\omega_n^2-\gamma_n^2/4}$. Using the retarded Green's function, the general solution for the cavity variable $q_n(t)$ follows as 
\begin{align}
 q_n(t) =& q_n^\mathrm{hom}(t;-\infty) + \int_ {-\infty}^t \mathrm{d}t'\ G_n(t-t') \vec{g}_n\cdot \vec{S}(t')   
\end{align}
where the homogeneous solution is exponentially decaying, i.e.,
$q_n^\mathrm{hom}(t;t') \propto e^{-\frac{\gamma_n (t-t')}{2}}$. Thus, $q_n^\mathrm{hom}(t;-\infty)=0$ for nonzero damping.  

Inserting $q_n(t)$ into the spin \acl{EOM} obtained from Eq.~\eqref{eq:H_toymodel}, $\dot{\vec{S}}(t)=-\vec{S}(t)\times (\vec{H}_0(t)+\sum_n \vec{g}_n q_n(t))$, gives
\begin{align}
    \begin{split}
    \dot{\vec{S}}(t)
    &=
    -\vec{S}(t)
    \hspace{-0.1em}
    \times 
    \hspace{-0.1em}
    \bigg[
    \vec{H}_0(t)
    \hspace{-0.1em}
    + 
    \hspace{-0.2em}
    \int_ {-\infty}^t
    \hspace{-0.75em}\mathrm{d}t'
    \sum_n 
    \hspace{-0.2em}
    G_n(t-t') 
    \vec{g}_n 
    \vec{g}_n^\mathrm{T}
    \vec{S}(t')
    \bigg].
    \end{split}
\end{align}
The memory kernel $\mathcal{K}(t-t') = -\sum_n G_n(t-t')\vec{g}_n\vec{g}_n^\mathrm{T}$ retains the full oscillator dynamics, so 
the spin-only \ac{nM-LLG} has the same eigenmode spectrum as the full coupled system.

Using the derivative expansion of the memory kernel and comparing with the \ac{iLLG}~\eqref{eq:iLLG}, we obtain the Gilbert damping and inertia parameter arising from coupling to the cavity modes
\begin{align}
    \alpha
    &
    =
    -
    \int_{0}^{\infty}
    \mathrm{d}\tau\
    \tau
    \mathcal{K}(\tau)
    =
    \sum_n
    \vec{g}_n\vec{g}_n^\mathrm{T}
    \frac{\gamma_n}{\omega_n^4},
    \\
    \eta
    &=
    \frac{1}{2}
    \int_{0}^{\infty}
    \mathrm{d}\tau\
    \tau^2
    \mathcal{K}(\tau)
    =
    \sum_n
    \vec{g}_n\vec{g}_n^\mathrm{T}
    \left(
    \frac{1}{\omega_n^4}
    -
    \frac{\gamma_n^2}{\omega_n^6}
    \right).
\end{align}
Gilbert damping requires cavity losses; the inertia parameter $\eta$, by contrast, is finite even in the lossless limit.

\textit{Eigenmode frequency spectrum}---We now compare the eigenmode spectra of the \ac{nM-LLG} and the \ac{iLLG} for this model. Assuming a constant $\vec{H}_0(t)=H_0\vec{e}_z$ and keeping terms up to first order in deviations from the ground state -- i.e., $\vec{S}(t)= \vec{S}_0+\vec{s}(t)$ with $|\vec{S}_0|\approx 1$, $\vec{S}_0\perp\vec{s}$, and $|\vec{s}|\ll 1$ -- we obtain a linearized \ac{nM-LLG}
\begin{align}
    \begin{split}
    \dot{\vec{s}}(t)
    \hspace*{-0.15em}
    &\approx
    \hspace*{-0.15em}
    -
    \Big(
    \vec{s}(t)
    \hspace*{-0.15em}
     \times 
     \hspace*{-0.15em}
     \tilde{\vec{H}}_0
    +
    \vec{S}_0
    \hspace*{-0.15em}
    \times 
    \hspace*{-0.15em}
    \hspace*{-0.25em}
    \int_ {-\infty}^t 
    \hspace*{-0.8em}
    \mathrm{d}t'
    \sum_n 
    G_n(t-t') 
    \vec{g}_n 
    \vec{g}_n^\mathrm{T}
    \vec{s}(t')
    \Big),
\end{split}
\end{align}
where the ground state fulfills $\vec{S}_0 \times \tilde{\vec{H}}_0
= 0$, with $ \tilde{\vec{H}}_0 = \vec{H}_0 -\kappa^{(0)}\vec{S}_0$.
Assuming harmonic time-dependence $\vec{s}(t)=\vec{s}e^{i\omega t}$, we get the linearized \ac{nM-LLG} in the frequency domain,
\begin{align}
    \begin{split}
    i
    \omega
    \vec{s}
    &\approx
    -
    \Big(
    \vec{s}
    \times 
    \tilde{\vec{H}}_0
    +
    \vec{S}_0
    \times 
    \sum_n 
    G_n(\omega)
    \vec{g}_n 
    \vec{g}_n^\mathrm{T}
    \vec{s}
    \Big),
\end{split}
\end{align}
with the Fourier transform of the retarded Green's function $G_n(\omega)=\int_0^{\infty} \mathrm{d}\tau\ e^{-i\omega \tau}G_n(\tau) =(\omega_n^2-\omega^2+i\gamma_n\omega)^{-1}$.

Next, we assume that there are three degenerate cavity modes with frequency $\omega_\mathrm{c}$, vanishing damping $\gamma_\mathrm{c}\rightarrow 0^+$, and orthogonal coupling vectors such that $\sum_{n=1}^3\vec{g}_\mathrm{n}\vec{g}_\mathrm{n}^\mathrm{T}=g_\mathrm{c}^2 \mathbbm{1}_{3\times3}$. The frequency spectrum can then be obtained by solving the two-dimensional secular equation
\begin{align}
    \mathrm{det}
    \left[
    \omega
    \mathbbm{1}_{2\times2}
    +
    \sigma_2
    \omega_0
    +
    \sigma_2
    \Sigma(\omega)
    \right]    
    &=0,
\end{align}
with the second Pauli matrix $\sigma_2=\left(\begin{smallmatrix} 0 & -i \\ i & 0 \end{smallmatrix}\right)$, and where we defined the bare magnon frequency $\omega_0=H_0$ and the magnon self-energy $\Sigma(\omega)=  \frac{g_\mathrm{c}^2}{\omega_\mathrm{c}^2}-\frac{g_\mathrm{c}^2}{\omega_\mathrm{c}^2-\omega^2}$. This equation has the implicit solution
\begin{align}
    \label{eq:omega_exact}
    \omega_\mathrm{exact}
    =
    \pm
    \omega_0
    \pm
    \Sigma(\omega_\mathrm{exact}),
\end{align}
which contains the exact (in the linearized limit) eigenmode frequencies of the coupled spin--cavity system.

Applying the same procedure to the \ac{iLLG} yields
\begin{align}
    \omega_\mathrm{iLLG}
    =
    \pm
    \omega_0
    \mp
    \omega^2_\mathrm{iLLG}
    \frac{g_\mathrm{c}^2}{\omega_\mathrm{c}^4}.
\end{align}
This is precisely what one obtains by expanding $\Sigma(\omega)$ in Eq.~\eqref{eq:omega_exact} to second order in $\omega$ -- the frequency-domain counterpart of the second-order time-derivative truncation that defines the \ac{iLLG}.

Unlike the implicit exact solution~\eqref{eq:omega_exact}, the \ac{iLLG} secular equation yields a closed-form expression for the frequencies,
\begin{align}
    \omega_\mathrm{iLLG}
    =
    -
    \xi 
    \frac{\omega_\mathrm{c}^4}{2g_\mathrm{c}^2}
    \pm 
    \sqrt{
    \frac{\omega_\mathrm{c}^8}{4g_\mathrm{c}^4}
    +
    \frac{\omega_\mathrm{c}^4}{g_\mathrm{c}^2}
    \omega_0},
\end{align}
with $\xi=\pm1$. The two positive-energy solutions are typically called precessional and nutational magnon modes~\cite{Mondal2023}.

In the limit of vanishing coupling, the precessional mode is simply given by
\begin{align}
    \lim_{g_\mathrm{c}\rightarrow 0}
    \omega^\mathrm{prec}_\mathrm{iLLG}
    =
    \omega_0,
\end{align}
while the nutational mode diverges as $1/g_\mathrm{c}^2$
\begin{align}
    \lim_{g_\mathrm{c}\rightarrow 0}
    \omega^\mathrm{nut}_\mathrm{iLLG}
    =
    \lim_{g_\mathrm{c}\rightarrow 0}
    \frac{\omega_\mathrm{c}^4}{g_\mathrm{c}^2}
    =
    \infty,
\end{align}
leaving no physical counterpart in the decoupled limit.

\begin{figure}
    \centering
    \includegraphics[width=1.0\linewidth]{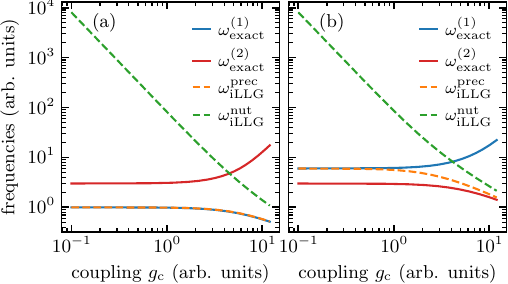}
    \caption{Comparison between the frequency spectrum obtained from the \ac{nM-LLG} and \ac{iLLG} for the bare magnon energy being lower (a) and higher (b) than that of the cavity modes.}
    \label{fig:comparison}
\end{figure}

The exact spectrum, by contrast, yields two well-defined modes in this limit:
\begin{align}
    \lim_{g_\mathrm{c}\rightarrow 0}
    \omega_\mathrm{exact}^{(1)}
    =
    \omega_0,
    ~~~ \textrm{and} ~~
    \lim_{g_\mathrm{c}\rightarrow 0}
    \omega_\mathrm{exact}^{(2)}
    =
    \omega_\mathrm{c} .
\end{align}
The frequencies obtained from the linearized \ac{nM-LLG} and \ac{iLLG} as a function of $g_\mathrm{c}$ are shown in Fig.~\ref{fig:comparison}. When the bare magnon frequency $\omega_0$ lies below the bare cavity frequency $\omega_\mathrm{c}$ [Fig.~\ref{fig:comparison}(a)], the low-frequency expansion underlying the \ac{iLLG} accurately describes the lower-energy mode, while its high-frequency root deviates strongly from the physical upper mode defined by $\omega_\mathrm{exact}^{(2)}$. The latter approaches $\omega_\mathrm{c}$ for $g_\mathrm{c}\rightarrow0$, whereas the \ac{iLLG} nutation root diverges as $g_\mathrm{c}^{-2}$. The exact \ac{nM-LLG} retains the cavity pole and correctly captures the resulting mode hybridization and level repulsion.

For a bare magnon frequency $\omega_0$ above the cavity resonance $\omega_\mathrm{c}$, even the low-energy \ac{iLLG} mode deviates from the exact result [Fig.~\ref{fig:comparison}(b)]. This reflects the pole structure of the self-energy, whose sign changes across $\omega_\mathrm{c}$ -- a feature its Taylor expansion around $\omega=0$ cannot reproduce. The \ac{iLLG} therefore loses validity as the relevant frequency approaches or exceeds the cavity resonance.

These results expose an important limitation of the \ac{iLLG}. The exact self-energy, $\Sigma(\omega)$, is replaced by its second-order low-frequency expansion, $\Sigma(\omega)\approx\Sigma(0)+\Sigma'(0)\omega+\frac{1}{2}\Sigma''(0)\omega^2$. This truncation changes the analytic structure of the response and generates a high-frequency nutation root. In particular, $\omega_\mathrm{iLLG}^{\mathrm{nut}} \approx \omega_\mathrm{c}^4/g_\mathrm{c}^2\gg\omega_\mathrm{c}$ for weak coupling, placing this root far outside the low-frequency regime $|\omega|/\omega_\mathrm{c}\ll1$ in which the expansion is valid. Thus, in the minimal model considered here, where spin inertia originates from coupling to additional dynamical degrees of freedom, the high-frequency \ac{iLLG} nutation root is a spurious, truncation-induced artifact rather than an independent pole of the exact coupled system.

\textit{Discussion}---Our minimal spin--cavity model demonstrates that the same bath coupling can give rise to a physically meaningful inertial response while the high-frequency nutation root generated by the \ac{iLLG} truncation is unphysical. When inertia arises from coupling to additional dynamical degrees of freedom, the inertial term represents the low-frequency expansion of the non-Markovian memory kernel. This expansion can accurately describe the low-frequency spin mode when the relevant bath frequencies lie well above the spin frequency, but it does not, by itself, imply an additional high-frequency spin excitation. A physical nutational mode requires a corresponding pole in the full response, which is not guaranteed by the truncated \ac{iLLG}.

This distinction between spin inertia and nutation extends beyond the cavity model. Several microscopic approaches derive inertial spin dynamics by eliminating additional dynamical degrees of freedom, resulting in a frequency-dependent spin response that is subsequently expanded in powers of frequency. This includes relativistic approaches based on the Dirac--Kohn--Sham equation, where eliminating the negative-energy components of the four-component spinor via the Foldy--Wouthuysen transformation generates an infinite series of time derivatives~\cite{Mondal2017,Mondal2018}, as well as descriptions of spins coupled to fermionic~\cite{Bhattacharjee2012,Kikuchi2015,Sayad2016,Bajpai2019,Faehnle2011,Thonig2017,Reyes-Osorio2025,Moussa2026} and bosonic~\cite{Anders2022,Quarenta2024} degrees of freedom. In the latter cases, the exact response retains the frequency-dependent structure associated with the eliminated degrees of freedom, including their characteristic poles, whereas the \ac{iLLG} retains only their low-frequency expansion. As in the model considered here, this truncation has the potential to generate spurious high-frequency roots; whether this occurs in a given microscopic derivation must be assessed from its full, untruncated response. The phenomenological models of Refs.~\cite{Ciornei2011,Wegrowe2012} likewise introduce an additional dynamical variable whose elimination leads to an inertial equation, albeit without a microscopic derivation.

\mw{That truncating the memory expansion leads to unphysical results is further solidified by retaining third-order time derivatives in Eq.~\eqref{eq:nM_expansion}, equivalent to expanding the self-energy in Eq.~\eqref{eq:omega_exact} to third order in frequency. The additional order introduces one further root, which for $\gamma_\mathrm{c}\ll g_\mathrm{c}^2/\omega_\mathrm{c}^2$ is found to be $\omega\approx -i\omega_\mathrm{c}^2/(2\gamma_\mathrm{c})$. Inserting this into the assumed harmonic time dependence gives $\vec{s}(t)\propto e^{\omega_\mathrm{c}^2/(2 \gamma_\mathrm{c}) t}$, a diverging and thus unphysical solution, whose growth rate moreover increases as the third-order coefficient $\kappa^{(3)}$ is reduced. This is directly analogous to the runaway solutions of the Abraham--Lorentz equation, which arise when the retarded self-interaction of a radiating charge is expanded in time derivatives \cite{Dirac1938}. Comparable artifacts occur in a broad range of reduced descriptions: Burnett-order truncations of the Chapman--Enskog expansion produce unstable short-wavelength modes absent from the exact spectrum of Boltzmann's equation \cite{Bobylev1982}, while in open quantum systems, eliminating the bath and truncating the resulting memory kernel to a time-local form yields the Redfield equation that can lead to negative populations \cite{Hartmann2020}. In each case the pathology arises outside the regime where the truncation is justified, and is removed by restoring part of the eliminated dynamics.}

Our analysis has direct experimental implications -- not for the existence of observed high-frequency features, but for their interpretation. A high-frequency spectral peak cannot, by itself, establish the presence of a nutational magnon. Existing assignments~\cite{Neeraj2021,Unikandanunni2022,De2025} identify such features by fitting the \ac{iLLG} and determining the inertia parameter from the peak frequency. However, agreement with the \ac{iLLG} frequency does not establish the microscopic origin of the peak: an additional dynamical degree of freedom coupled to the spins can produce a high-frequency feature at a comparable frequency through mode hybridization. A microscopic identification of nutation therefore requires evidence for a corresponding pole in the full response, rather than a fit to the peak position alone.

\begin{figure}[t]
    \centering
    \includegraphics[width=1.0\linewidth]{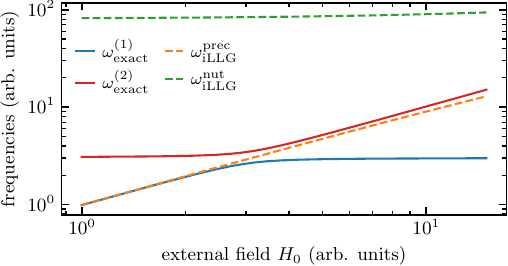}
    \caption{{\blue Hybridization of a magnon and a bath mode.} The linearized \ac{nM-LLG} captures the avoided crossing between magnon and bath {\blue modes} at the resonance between the bare modes. The character of the modes reverses at the avoided crossing. The linearized \ac{iLLG} only reproduces the behavior of the low-energy magnetic mode far away from the resonant regime.
    }
    \label{fig:comparison_2}
\end{figure}

Resonant hybridization provides a well-established example of the failure of Markovian descriptions. When a bath mode approaches resonance with a magnon, the frequency dependence of the memory kernel becomes essential and cannot be captured by its second-order low-frequency expansion. {\blue As illustrated in Fig.\ \ref{fig:comparison_2},} the full response then exhibits an avoided crossing, across which the character of the two modes evolves from predominantly magnonic to predominantly bath-mode like and vice versa. 
Far from resonance, the \ac{iLLG} can still reproduce the low-energy magnetic mode, but it cannot describe this resonant hybridization {\blue [Fig.~\ref{fig:comparison_2}]}. Observed magnon--phonon anticrossings~\cite{Hay1970,Jensen1975} thus illustrate why the dynamical degrees of freedom responsible for the memory kernel must be retained when their characteristic frequencies enter the relevant spectral range.

These considerations also place recent experimental observations in a broader context. Terahertz spectroscopy of ferromagnetic cobalt found {\blue additional high-frequency spectral features} that {\blue could be explained by} non-Markovian coupling to a phonon bath, {\blue i.e.,}
beyond those described by the \ac{iLLG} nutation mode~\cite{Hartmann2025}. Such observations underscore the need to distinguish high-frequency features arising from the dynamical degrees of freedom underlying the memory kernel from those associated with a genuine nutational pole of the full spin response.

{\blue Lastly,} 
our model 
suggests a direct experimental discriminator. Since $\eta=g_\mathrm{c}^2/\omega_\mathrm{c}^4$, reducing the coupling {\blue strength} $g_\mathrm{c}$ lowers the inferred inertia and drives the \ac{iLLG} nutation frequency $\omega^\mathrm{nut}_\mathrm{iLLG}\approx\omega_\mathrm{c}^4/g_\mathrm{c}^2$ to higher frequencies. The corresponding upper mode of the full spin--cavity system, in contrast, remains finite and approaches $\omega_\mathrm{c}$ as $g_\mathrm{c}\rightarrow0$. Varying the coupling to the degrees of freedom responsible for the memory kernel, for example through the coupling geometry~\cite{Huebl2013,Tabuchi2014}, could therefore {\blue lead to distinguishing} 
a truncation-induced \ac{iLLG} root from a genuine hybridized mode.

\mw{
\textit{Conclusion}---Identifying high-frequency spin excitations requires the complete, non-Markovian memory kernel. In the minimal spin--cavity model considered here, the nutation branch of the \ac{iLLG} arises from the second-order truncation of that kernel and does not correspond to an independent mode of the full system: it is a root generated beyond the validity range of the expansion it is obtained from. Finite-order reductions of coupled dynamical systems can therefore produce spurious spectral features. Retaining the full memory kernel, by contrast, keeps the poles of the eliminated degrees of freedom and with them the hybridization physics -- level repulsion and avoided crossings -- that the \ac{iLLG} cannot represent. {\blue We propose that} coupling-dependent measurements, interpreted against the full frequency dependence of the response, could then distinguish genuinely physical excitations from truncation artifacts, and resolve the character of high-frequency magnetic excitations directly.
}

\textit{Acknowledgments.}---M.W. thanks Felix Hartmann for stimulating discussions.  P.M.O.\ acknowledges funding from the German Research Foundation (Deutsche Forschungsgemeinschaft) through CRC/TRR 227 “Ultrafast Spin Dynamics” (Project MF, Project ID No.\ 328545488). This work was further supported by the Swedish Research Council (VR) and the Knut and Alice Wallenberg Foundation (Grants No.\ 2022.0079 and No.\ 2023.0336). We acknowledge funding from the European Union’s HORIZON EUROPE, under Grant Agreement No.\ 101129641, “OBELIX”.

%

\end{document}